\documentclass[aps,amsmath,twocolumn,amssymb,floatfix,showpacs,superscriptaddress,nofootinbib,longbibliography]{revtex4-1}
\usepackage{mathtools}
\usepackage{braket}
\usepackage[dvipsnames]{xcolor}
\usepackage{float}
\usepackage{subfigure}
\usepackage{graphics}
\usepackage{bm}
\usepackage{soul}
\usepackage{MnSymbol}
\usepackage{braket}
\usepackage{float}
\usepackage{tikz}
\usepackage{blkarray}
\usepackage{pgffor}
\usepackage{float}
\usepackage{multirow}
\usepackage{booktabs}
\usepackage{silence}
\usepackage{makecell}

\usepackage[colorlinks=true,linktoc=page,linkcolor=BrickRed,citecolor=magenta,urlcolor=purple]{hyperref}

\mathchardef\mhyphen="2D 

\newcommand\bea{\begin{eqnarray}}
\newcommand\eea{\end{eqnarray}}
\newcommand\beq{\begin{equation}}  
\newcommand\eeq{\end{equation}}

\usepackage[normalem]{ulem}
\definecolor{lime}{HTML}{A6CE39}
\usepackage{sidecap,tikz}
\DeclareRobustCommand{\orcidicon}{\hspace{-1.0mm}
	\begin{tikzpicture}
		\draw[lime, fill=lime] (0.0,0.0) 
		circle [radius=0.15] 
		node[white] {{\fontfamily{qag}\selectfont \tiny \,ID}};
		\draw[white, fill=white] (-0.0525,0.095) 
		circle [radius=0.007];
	\end{tikzpicture}
	\hspace{-3.0mm}
}
\foreach \x in {A, ..., Z}{\expandafter\xdef\csname orcid\x\endcsname{\noexpand\href{https://orcid.org/\csname orcidauthor\x\endcsname}
		{\noexpand\orcidicon}}
}

\AtBeginDocument{%
	\newwrite\bibnotes
	\def\bibnotesext{Notes.bib}
	\immediate\openout\bibnotes=\jobname\bibnotesext
	\immediate\write\bibnotes{@CONTROL{REVTEX41Control}}
	\immediate\write\bibnotes{@CONTROL{%
			apsrev41Control,author="08",editor="1",pages="1",title="1",year="1"}}
	\if@filesw
	\immediate\write\@auxout{\string\citation{apsrev41Control}}%
	\fi
}%
\usepackage{orcidlink}
\begin{document}

\title{Altermagnetic phases and phase transitions in Lieb-$5$ Hubbard model}
\author{Sougata Biswas\,\orcidlink{0000-0002-9158-3904}}
\email{sougatabiswas@prl.res.in}
\affiliation{Theoretical Physics Division, Physical Research Laboratory, Navrangpura, Ahmedabad, Gujarat-380009, India}
\author{Achintyaa\,\orcidlink{0009-0006-7739-007X}}
\email{achintyaa@prl.res.in}
\affiliation{Theoretical Physics Division, Physical Research Laboratory, Navrangpura, Ahmedabad, Gujarat-380009, India}
\affiliation{Indian Institute of Technology Gandhinagar, Palaj, Gujarat-382055, India}
\author{Paramita Dutta\,\orcidlink{0000-0003-2590-6231}}
\email{paramita@prl.res.in}
\affiliation{Theoretical Physics Division, Physical Research Laboratory, Navrangpura, Ahmedabad, Gujarat-380009, India}

\vspace{0.8cm}
\begin{abstract}
The emergence of altermagnetism, the collinear magnetic phase characterized by momentum-dependent spin-split bands but zero net magnetization, has fundamentally reshaped the classification of magnetic order. We propose an altermagnetic (AM) order in a repulsive Hubbard model on the Lieb-$5$ lattice. Considering only nearest-neighbor hoppings within the lattice, we show a phase transition from the nonmagnetic to a unique AM isolated band metal phase (AMIM), allowing clear identification of spin-split states. Additionally, the AM metallic phase (AMM) is also shown to appear as an intermediate phase during the transition from the normal metal to the AMIM in the presence of the diagonal hopping within each unit cell of the Lieb-$5$ lattice. The manifestation of distinct AM phases and the phase transitions, driven by Hubbard interaction and hopping integrals, have been explored in terms of spin-resolved band structure, spectral function, and the behavior of the AM order parameter. The stability of these AM phases against the spin-orbit coupling and temperature is also established. 
\end{abstract}
\maketitle

\section{Introduction}
Magnetic materials are traditionally classified into two primary types based on the spin ordering. Ferromagnets (FM) host spin-split energy bands with a finite magnetic moment originating from the parallel alignment of spins. Antiferromagnets (AFM) in which neighboring spins align antiparallel, possess zero net magnetization, and exhibit spin-degenerate energy bands. Beyond this traditional classification of magnetism, a new class of two-dimensional ($2$D) magnetic order, termed as `altermagnetism', has been recently identified~\cite{Hayami2019,Smejkal2020,Yuan2020, Hayami2020, Smejkal2022a, Smejkal2022b, Mazin2021,Feng2022, Mazin2022,Ifmmode2022c,Gonzalez2023, Gonzalez2024, Krempasky2024, Mazin2024, McClarty2024,Song2025}. The momentum-dependent spin-splitting of the energy bands together with the zero net magnetization property makes the altermagnet (AM) to partially share characteristics of both FM and AFM, yet remain fundamentally distinct from either~\cite{Smejkal2022b,Smejkal2022a,Mazin2022,McClarty2024,Song2025}. The cancellation of the net magnetization in AMs arises from the alternating spin orientation on neighboring two sublattices with broken Kramer’s spin-degeneracy~\cite{Krempasky2024, Lee2024}. However, unlike in conventional AFM, these opposite spin sublattices follow a specific crystal rotation that is inherent to the lattice symmetry~\cite{Das2024, Che2025}. 

Owing to the unique combination of FM-like spin splitting and AFM-like zero net magnetization, AMs support strong spin currents, high-frequency dynamics with minimum stray field~\cite{Giil2024, Baltz2024}. AM phases have been realized experimentally in a wide range of materials starting from RuO$_2$~\cite{Fedchenko2024,Zhou2024, Guo2024a,Guo2024b,Lin2024}, and extended to CrSb~\cite{Reimers2024, Yang2025, Ding2024, Li2025}, K-Cl~\cite{Naka2019, Naka2020, Yu2025}, MnTe~\cite{Mazin2023, Krempasky2024, Osumi2024, Lee2024}, MnF$_2$~\cite{Yuan2020, Bhowal2024}, FeSb$_2$~\cite{Mazin2021}, GdAlSi~\cite{Nag2024}, KV$_2$Se$_2\text{O}$~\cite{Jiang2025}, and others~\cite{Roig2024, Guo2023}. They have emerged as promising materials, exhibiting a wide range of intriguing physical properties including superconductivity~\cite{Zhang2024a,Zhao2025a,Ouassou2023,Chakraborty2024,Sim2025,Hong2025,Debnath2025,Sukhachov2024,Papaj2023,Zhu2023,Li2023,Ghorashi2024,Li2024b,Beenakker2023,Cheng2024a,Zyuzin2024,Lu2024,Bose2024,Maeland2024,Hu2025, Li2026}, magnetoresistance~\cite{Ifmmode2022c,Gonzalez2024,Chi2024,Liu2024}, topological phases~\cite{Del2025,Li2024a,Ghorashi2024} with potential of applications in spintronics~\cite{Bai2024,Sourounis2025,Gonzalez2021,Shao2021, Naka2021,Zhang2024d} and other fields~\cite{Song2025, Guo2024b,Zhang2024b,Guo2024c, Hariki2024,Gonzalez2023,Rathore2025,Yan2024,Fender2025}.

Altermagnetic phases have been investigated across a variety of lattice models: square lattice~\cite{Das2024,He2025,Consoli2025,Che2025,Maier2023}, Shastry-Sutherland lattice~\cite{Ferrari2024}, square-octagon lattice~\cite{Bose2024, Zhu2025, Vijayvargia2025, Che2025}, hexa-triangular lattice~\cite{Consoli2025}, honeycomb lattice~\cite{Zhu2025, Sato2024}, Kondo lattice~\cite{Zhao2025a}, Kagome lattice~\cite{peces2025}, and others~\cite{Zhu2025, Che2025, Giuli2025},
demonstrating how crystal symmetries combined with appropriate spin distribution can give rise to alternating spin polarization and spin-split band structures without net magnetization in various ways. Depending on the local environment and the strength of the interaction, these lattices have been found to exhibit altermagnetic-metallic (AMM) or an altermagnetic-insulating (AMI) phase~\cite{Das2024,He2025}. Among various lattices, the traditional Lieb lattice, often regarded as a prototype for oxychalcogenides or anti-CuO$_2$ materials, has drawn attention because of the simplicity of the lattice being very close to the square lattice model~\cite{Kaushal2025,Durrnagel2025,Antonenko2025}. 
\begin{figure}
    \centering
\includegraphics[width=.95\linewidth]{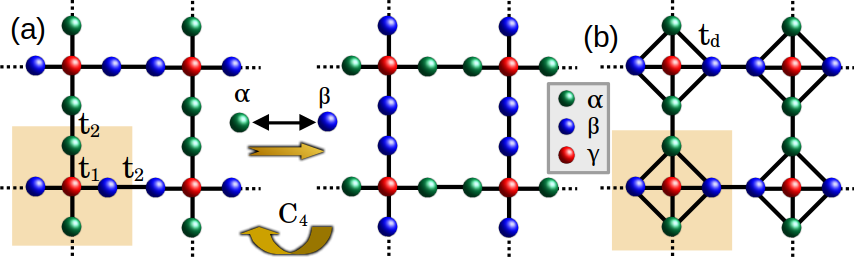}
\caption{(a) Lieb-$5$ lattice containing three sub-lattices invariant under a spin flip followed by a $C_4$ rotation around the center. (b) Its variant with diagonal hopping.}
\label{lattice_figure}
\end{figure}
The Lieb lattice, characterized by various electronic and topological features arising from its extended structural connectivity, has served as an important model for exploring various physical phenomena. These include localization-delocalization properties~\cite{Liu2020, Lima2025, Liu2021, Zhang2019, Li2022}, topological phases and edge states~\cite{Bhattacharya2019, Weeks2010, Goldman2011, Palumbo2015, Kunst2019, Li2018, Banerjee2021}, magnetic phases~\cite{Noda2014, Costa2016, Noda2015, Nikolaenko2025, Nie2017}, and superconducting phases~\cite{Swain2020, Penttila2025, Yamazaki2020}. Lieb lattices have been realized experimentally~\cite{Roman2021, Vicencio2015, Slot2017, Centala2023, Whittaker2018, Che2025, Cui2020, Feng2020, Diebel2016} showcasing the potential of application across various fields~\cite{Roman2021, Vicencio2015, Lazarides2017, Ablowitz2019, Scafirimuto2021, Zhu2018, Yang2016, Casteels2016}. The AM phase in this lattice has been shown to arise from the coexistence of magnetic and non-magnetic sites at specific Wyckoff positions~\cite{Kaushal2025,Durrnagel2025}.

In this article, {\it we demonstrate a type of AM phase, altermagnetic isolated band metal (AMIM) phase and also AMM phase, and their transitions in Lieb-$5$ lattice~\cite{Liu2020, Liu2021, Bhattacharya2019, Norian2025}} based on a repulsive Hubbard model. In contrast to the AMM phase found in the traditional Lieb (also known as Lieb-$3$) metal~\cite{Durrnagel2025, Che2025, Kaushal2025}, {\it we show unique AM phase in the Lieb-$5$ Hubbard model without any diagonal hopping}. We establish the AM phase in our model by investigating the behavior of the spin-resolved band structure and spectral density function. Inclusion of next-nearest neighbor hopping integral allows us to further control the phase diagram. The effect of the intrinsic spin-orbit coupling has also been studied to explore the stability of the AM phases in the present lattice. The simplicity of our lattice model showing multiple AM phases, including the unique AMIM phase, even without any second-neighbor interaction term, enhances the importance of the present work. 

\section{Model and method} We consider a repulsive Hubbard model in a $2$D Lieb-$5$ lattice consisting of three sublattices, $\alpha$, $\beta$, and $\gamma$, within a unit cell as shown in Fig.~\ref{lattice_figure} (yellow-shaded area). The Hamiltonian reads,
\begin{equation}
  \mathcal{H} = - \sum_{i,j, \sigma} t_{ij} (c_{i \sigma}^\dagger c_{j \sigma} + h.c.) + U \sum_{i} n_{i \uparrow}n_{i \downarrow} + \varepsilon \sum_{i,\sigma} n_{i\sigma}, 
\label{eq:Hamiltonian}
\end{equation}
where $c_{i,\sigma}^{\dagger}$ $(c_{i,\sigma})$ operator creates (annihilates) an electron of spin $\sigma$ ($=\uparrow,\downarrow$) at $i$-${\rm th}$ site. The hopping integral $t_{ij}$ can take values $t_1$ and $t_2$ for the intracell and intercell hopping, respectively, whereas the diagonal hopping is denoted by $t_d$. The term $\varepsilon$ is the onsite potential fixed to zero for all sublattices throughout our calculations. The repulsive Hubbard interaction is introduced through the second term of Eq.~\ref{eq:Hamiltonian} with $U$ as the on-site interaction strength. The system can return to its original configuration when two consecutive transformations $(i)$ spin flip between $\alpha$ and $\beta$ sublattices (exchanging green and blue sites) and $(ii)$ $[C_4]$ rotation about the $\gamma$ site are applied. 

In order to investigate the magnetic order, we carry out a self-consistent Hartree–Fock (HF) analysis which captures the onset of magnetic phases decided by the underlying lattice geometry and interaction strength. The AM order parameter ($\delta m$)  proportional to the staggered magnetization can be defined as~\cite{Das2024}
\begin{equation}
     \delta m  =  \frac{1}{8N}  \langle \sum_{i \in \alpha }(n_{i \uparrow} - n_{i \downarrow}) -\sum_{i \in \beta}(n_{i \uparrow}-n_{i \downarrow})\rangle_{HF}
\end{equation}
ensuring the finite contribution from the $\alpha$ and $\beta$ sublattices and no role of $\gamma$ sublattice as we consider non-zero $\delta m$ only at $\alpha$ and $\beta$ sublattices); $\text{N}$ being the total number of unit cells. A non-zero $\delta m$ with an alternating spin splitting essentially indicates an AM phase. Unless mentioned, throughout the present work, we consider half-filling and express the occupation number being sensitive to both the sublattice and the spin in the form,
\begin{equation}
    \langle n_{i\sigma}\rangle = \frac{1}{2}+
    \left\{
    \begin{array}{cc}
         +\delta m~ (-1)^\sigma & i\in \alpha \\
          -\delta m ~(-1)^\sigma& i\in \beta\\
          0& i\in \gamma
    \end{array}
    \right.
    \label{Eq:occupation}
\end{equation} 
where the spin index $\sigma$ is $0$ for $\uparrow$-spin and $1$ for $\downarrow$-spin. Within the mean-field approximation, the Hubbard term of the Hamiltonian simplifies as $ U \sum_{i}n_{i\uparrow} n_{i\downarrow}\approx \langle n_{i\uparrow}\rangle n_{i\downarrow}+ n_{i\uparrow}\langle n_{i\downarrow}\rangle-\langle n_{i\uparrow}\rangle\langle n_{i\downarrow}\rangle$ which can be decoupled as $-U\delta m\left[\sum_{i\in \alpha}( n_{i\uparrow}- n_{i\downarrow})-\sum_{i\in \beta}( n_{i\uparrow}- n_{i\downarrow})\right] + c$  using Eq.~\eqref{Eq:occupation} where $c$ is a constant term. \footnote{The constant term $c$ = $\frac{U}{2} \sum\limits_{i \in \alpha, \beta, \gamma} (n_{i\uparrow}+n_{i\downarrow})$ - $2U (\frac{1}{4}-\delta m^2)$ - $\frac{U}{4}$ provides an energy shift and does not affect our results qualitatively~\cite{Das2024}. We neglect it and always measure the energy with respect to the chemical potential.}
The mean-field Hamiltonian in the momentum space is taken as,
\begin{equation}
\begin{split}
  H  = \sum_{k}\Psi_{k}^\dagger \vcenter{\hbox{$\begin{bmatrix}
    H_{\uparrow}(k) & 0 \\
    0 & H_{\downarrow}(k)\\
\end{bmatrix}$}} \Psi_{k}   
\end{split}
\label{Eq:MFHamiltonian}
\end{equation}
where,
\begin{equation}
{H_{\uparrow(\downarrow)}}(\mathbf{k})=
\left(\def\arraystretch{1.5} \begin{matrix}
\epsilon_{\alpha_{\uparrow(\downarrow)}} & -t_d & -t_2 e^{ik_{y}} & -t_d & -t_1  \\
-t_d & \epsilon_{\beta_{\uparrow(\downarrow)}} & -t_d & -t_2 e^{-ik_{x}} & -t_1 \\
-t_2 e^{-ik_{y}} & -t_d & \epsilon_{\alpha_{\uparrow(\downarrow)}} & -t_d & -t_1 \\
-t_d & -t_2 e^{ik_{x}} & -t_d & \epsilon_{\beta_{\uparrow(\downarrow)}} & -t_1  \\
-t_1 & -t_1 & -t_1 & -t_1 & \epsilon_{\gamma_{\uparrow(\downarrow)}} \\
\end{matrix}\right)
\label{Eq:Hamiltonian_matrix}
\end{equation}
considering the basis $\Psi_{k}^\dagger = (c_{{k} \alpha \uparrow}^\dagger, c_{{k} \beta \uparrow}^\dagger,c_{{k} \alpha \uparrow}^\dagger, c_{{k} \beta \uparrow}^\dagger, c_{{k} \gamma \uparrow}^\dagger,c_{{k} \alpha \downarrow}^\dagger, c_{{k} \beta \downarrow}^\dagger,c_{{k} \alpha \downarrow}^\dagger, c_{{k} \beta \downarrow}^\dagger, c_{{k} \gamma \downarrow}^\dagger)$. The onsite terms are basically, $\epsilon_{\alpha_{\uparrow,\downarrow}} = \mp U \delta m$, $\epsilon_{\beta_{\uparrow,\downarrow}} = \pm U \delta m$ and $\epsilon_{\gamma_{\uparrow,\downarrow}} = 0$. We take the Fourier transformation of the Hamiltonian and calculate the order parameter self-consistently with a tolerance of $10^{-5}$. Further details of the calculation can be found in Appendix~\ref{AppendixA}. 

\begin{figure}
\centering
\includegraphics[scale=0.3]{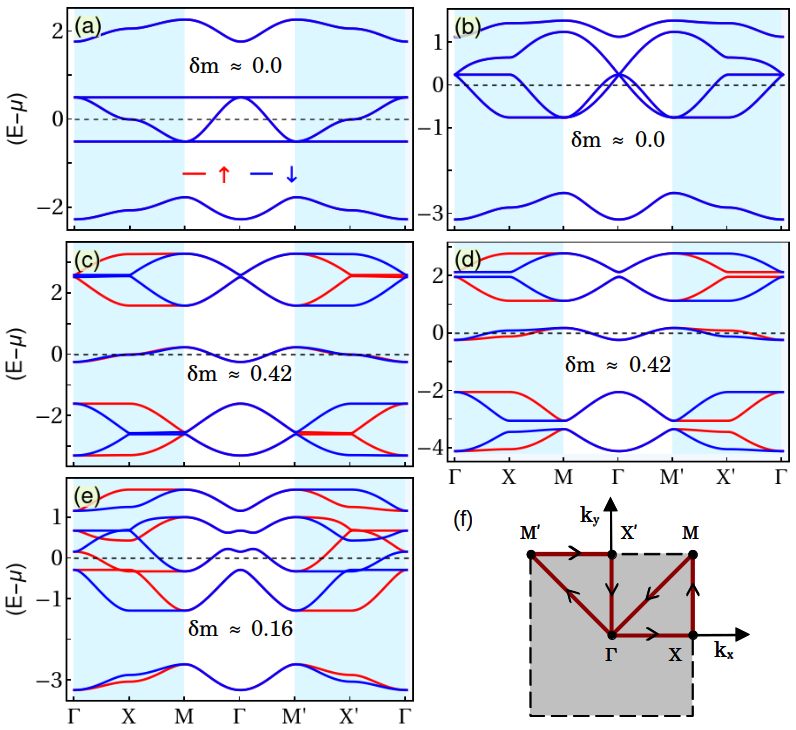}
\caption{Spin-resolved band structure of Lieb-$5$ lattice in (a,b) NM phase (a) without ($t_d=0$) and (b) with diagonal hopping ($t_d=0.5$), and (c-e) for AM phase (c) without ($t_d=0$) and (d,e) with diagonal hopping ($t_d=0.5$). The Hubbard interaction term is considered as (a,b) $U = 0$ (c,d) $U = 5$, and (e) $U = 3$. The other parameters are: $t_{1} = 1$, $t_{2} = 0.5$. (f) Schematic diagram of the first Brillouin zone (BZ) with high symmetry points and path ($k_x,k_y \in [-\pi, \pi]$). }
\label{fig:EK}
\end{figure}

\section{Results and Discussion} 
\subsection{Order parameter}
We start by demonstrating the band structure of the Lieb-$5$ lattice without any Hubbard term i.e., $U = 0$ as shown in Fig.~\ref{fig:EK}(a,b). There is no spin splitting with and without diagonal hopping. These correspond to the normal phase. In the presence of Hubbard interaction, the emergence of AM phase is uncovered in the model without any second neighbor hopping of Lieb-$5$ lattice through the spin-resolved band structure depicted in Fig.~\ref{fig:EK}(c). Then, we study the effect of the diagonal hopping integral on the AM phase in Fig.~\ref{fig:EK}(d-e). We plot the band structures along paths connecting the high symmetry points of the Brillouin zone (BZ) shown in  Fig.~\ref{fig:EK}(f): $\Gamma(0,0) 
\xrightarrow[k_y = 0]{}
X(\pi,0)
\xrightarrow[k_x = \pi]{}
M(\pi,\pi)
\xrightarrow[k_x = k_y]{}
\Gamma(0,0)
\xrightarrow[-k_x = k_y]{}
M'(-\pi,\pi)
\xrightarrow[k_y = \pi]{}
X'(0,\pi)
\xrightarrow[k_x = 0]{}
\Gamma(0,0)$. From Fig.~\ref{fig:EK}(c-e), it is evident that the energy bands are spin-split along $\Gamma$-$X$-$M$ path which encodes two paths parallel to $k_x$ and $k_y$-axis consecutively (see Fig.~\ref{fig:EK}(f)). In contrast, it shows spin-degeneracy along $M$-$\Gamma$ or equivalently $\Gamma$-$M'$ path (diagonal path). Comparing the band structure along $\Gamma$-$X$-$M$-$\Gamma$ with that for $\Gamma$-$M^{\prime}$-$X^{\prime}$-$\Gamma$ path, a momentum-inversion symmetry is unraveled, confirming the zero net magnetization in the system. This momentum-dependent spin-splitting, along with the zero net magnetization, indicates the AM phase in the system. It appears both with and without diagonal hopping. 

With the onset of the AM phase, we observe that the AM phase emerging in the Lieb-$5$ lattice is basically an isolated band metal type with a single band around the Fermi energy, maintaining an energy gap from other bands as seen in Fig.~\ref{fig:EK}(c). The appearance of this unique phase is completely different from AM phases found in Lieb metal~\cite{Durrnagel2025} and also in other lattices~\cite{Das2024,He2025,Consoli2025,Che2025,Maier2023, Ferrari2024, Bose2024, Zhu2025, Vijayvargia2025, Che2025, Consoli2025, Zhu2025, Sato2024, Zhu2025, Che2025, Giuli2025} so far in the literature. We call it altermagnetic isolated band metal (AMIM). Naturally, a question arises about the robustness of the AMIM phase upon the inclusion of the diagonal hopping $t_d$. As seen in Fig.~\ref{fig:EK}(d), the AMIM phase remains preserved in the presence of the diagonal hopping for the same Hubbard strength. However, the scenario is different when we tune the Hubbard strength. For lower $U$ as shown in Fig.~\ref{fig:EK}(e), while the AM phase is ensured, the AMM emerges instead of AMIM phase in the presence of the diagonal hopping integral. We get back the AMIM phase in the presence of the diagonal hopping for larger Hubbard strength, confirming the sensitivity of the AM phase to both the Hubbard interaction and hopping strength. Note that, from the symmetry perspectives, AMM and AMIM phases are equivalent but with different responses. In the AMM phase, multiple bands appear around the Fermi level, while in the AMIM, due to the presence of a single band pair at the Fermi level, the spin splitting is not obscured by multiband scatterings and can be probed by unique spin-polarized transport signatures. It denotes a distinct electronic configuration of the AM state characterized by the isolated bands around the Fermi level.
\begin{figure}
\centering
\includegraphics[scale=0.3]{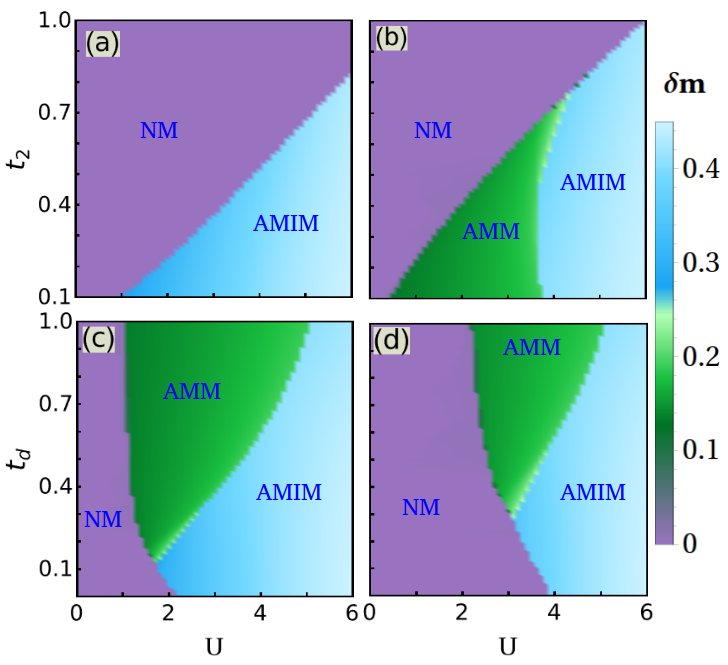}
\caption{(a,b) Density plot of $\delta m$ with Hubbard $U$ and (a-b) intercell hopping integral $t_2$ 
for (a) $t_d=0$, (b) $t_d=0.5$ and (c-d) diagonal hopping integral for (c) $t_2 =0.25$, (d) $t_2 = 0.5$. Other parameter values are the same as in Fig.~\ref{fig:EK}.}
\label{fig:Density}
\end{figure}

In order to investigate the interplay of the Hubbard strength and the hopping integral, we now analyze the behavior of the AM order parameter $\delta m$ with Hubbard strength $U$ and the intercell $t_2$ (diagonal hopping parameter $t_d$) in Fig.~\ref{fig:Density}(a-b) ((c-d)). The zero $\delta m$ corresponds to the NM phase (violet region), while the non-zero values of $\delta m$ indicate two different AM phases: (i) AMM phase with lower amplitudes of $\delta m$ (green region) and (ii) AMIM phase with higher values of $\delta m$ (blue region). We confirm the distinction between the two AM phases from the band structure. For a given $t_2$, there exists a threshold value of $U$ to obtain the AM phase, e.g., $U_{\text{th}} \sim 1$ for $t_2=0.1$. For the Hubbard strength lower than  $U_{\text{th}}$, the spin splitting completely disappears, resulting in a normal metallic (NM) phase with zero $\delta m$ (see Fig.~\ref{fig:Density}(a)). Thus, a \textit{phase transition from the NM to AMIM driven by the Hubbard interaction strength occurs in Lieb-$5$ lattice without any second-neighbor interaction}, while to get the AM phase in the traditional Lieb lattice in the second nearest neighbor hopping is essential~\cite{Durrnagel2025}. The reason behind the threshold $U_{\text{th}}$ lies in the underlying lattice geometry and the distribution of the order parameter $\pm U \delta m$ among the sublattices and spin configuration. They are proportional to the staggered magnetization, maintaining a distribution within each unit cell of the lattice geometry. When both the intra- and intercell hopping strengths ($t_1$ and $t_2$) are comparable, the mobility of the electrons within the lattice increases, bringing the system back to the NM phase for the same Hubbard strength. The staggered onsite configuration gets affected, and the system no longer supports the AM phase. Thus, preserving the AM phase in the system is possible by keeping $t_2$ always less than $t_1$. For a fixed $t_1$, the required $U_{\text{th}}$ to get the AM phase will be proportional to $t_2$. Note that, $U$ is constrained in the mean field approximation.

With the inclusion of the diagonal hopping $t_d$, another metallic phase along with a spin-split configuration, {\it the AMM phase emerges as an intermediate phase during the transition from the NM to AMIM phase} by tuning the Hubbard strength as shown in Fig.~\ref{fig:Density}(b). The intermediate AMM phase appears because the hopping $t_d$ allows an extra path for the electrons to hop between $\alpha$ and $\beta$ sublattices within each unit cell, enhancing the mobility of the electrons in the system. In fact, further increasing this intercell hopping integral brings the NM phase back to the system for the given $U$. The role of $t_d$ behind the appearance of the intermediate AMM phase can be understood more clearly when we plot the phase diagram as a function of $t_d$ keeping $t_2$ fixed (see Fig~\ref{fig:Density}(c-d)). For a given $t_2$, the phase transition from the NM to AMM to AMIM occurs, showing the AMM phase for a wide range of $U$ when $t_2$ is small. Increasing $t_2$ further will increase the threshold value of $t_d$ to get the AMM phase as seen by comparing Fig~\ref{fig:Density}(c) and (d)). Also, the AMM region of the phase diagram gets squeezed for higher $t_2$. Thus, by {\it tuning $t_2$ and $t_d$ properly, one can obtain an AMM phase over a large $U$ region in addition to the AMIM phase}. Thus, the role of the finite diagonal hopping is dual: (i) it lowers the threshold on $U$ for the emergence of the AM phase and (ii) it renders two consecutive phase transitions, NM to AMM to AMIM, driven by the Hubbard $U$.
\begin{figure}
\centering
\includegraphics[scale=0.32]{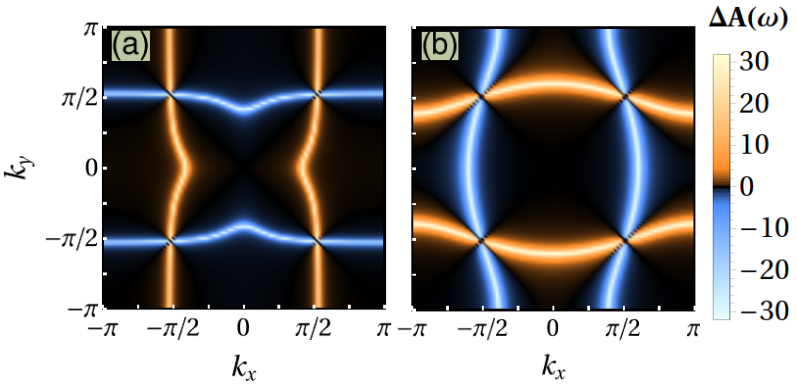}
\caption{Spin-resolved spectral function $\Delta A(\omega)$ = $A_{\uparrow}(\omega) - A_{\downarrow}(\omega)$ in $k_x, k_y$ space for Lieb-$5$ lattice with (a) $U =3$ and (b) $U = 5$ considering $t_d = 0.5$, $\omega = 0$ and $\eta = 0.01$. The other parameters are the same as in Fig.~\ref{fig:EK}.}
\label{fig:spectral}
\end{figure}
The inclusion of $t_d$ increases the hybridization or overlap between orbitals, which increases the possibility of band overlapping and subsequently, the density of states around the Fermi level. Due to the enhancement of the density of states around the Fermi level, a relatively smaller $U$ is sufficient to induce the AM phases. So, $t_d$ not only enables the emergence of the AMM phase but also reduces the critical interaction strength. Note that, we also calculate $\delta m$ in the real space as shown in the Appendix~\ref{AppendixA} and find an excellent agreement with the results in the momentum space. In Appendix~\ref{AppendixB}, we discuss the NM phase spin-resolved band structures in detail and also show the full band structures for all phases. We consider zero temperature throughout the main text. However, for the sake of completeness, we discuss the effect of the temperature and filling in Appendix~\ref{AppendixC} and discuss other traditional magnetic phases in Appendix~\ref{AppendixD}.

\subsection{Spectral function}
With an understanding of the band structure and phase diagram, we now investigate the spin-resolved spectral function, which is directly related to the angle-resolved photoemission spectroscopy (ARPES) spectra in experiments. It is defined as
\begin{equation}
A_{\uparrow(\downarrow)}(\omega) = \frac{\eta}{\pi} 
\sum_{k} 
\frac{|\langle k | \uparrow(\downarrow) \rangle|^{2}}
{(\omega - (E_{k}-\mu))^{2} + \eta^{2}},
\label{Eq:Spectral-function}
\end{equation}
where $\omega$ is the energy of the incoming electron with spin $\uparrow$ ($\downarrow$) and eigen state $|\uparrow \rangle$ ($|\downarrow \rangle$), $|k\rangle$ is the eigenstate of the system corresponding to the energy $E_k$, $\mu$ is the chemical potential, and $\eta$ is the spectral broadening factor.

In Fig~\ref{fig:spectral}, we plot the difference between the two spectral functions corresponding to the two spins at $\omega = 0$ in the momentum space for different Hubbard strengths in the presence of $t_d$. These parameter regimes correspond to the AMM and AMIM phases in Fig.~\ref{fig:EK}, respectively. The difference spectral function $\Delta A(\omega)$ exhibits alternating sign changes along different directions in the $k$-space with crossings at certain symmetry points following the $[C_4]$ symmetry (see Fig.~\ref{fig:spectral}(a)). The nature of alternating sign changes is also preserved in the AMIM phase as shown in Fig.~\ref{fig:spectral}(b), indicating a well-defined momentum-dependent spin polarization with $d$-wave symmetry, inherently favoring AM phases. Before we leave this section, the spin-resolved spectral functions for both up and down spins is also computed along the high-symmetry path for non-zero $\omega$ (see Appendix~\ref{AppendixB}).

\subsection{Spin-orbit interaction}
In order to examine the stability of the AM phases against the intrinsic spin–orbit coupling (SOC), we compute the spin-resolved band structure of the Lieb-$5$ lattice by incorporating SOC into the Hamiltonian as~\cite{Kane2005, Bhattacharya2019, Beugeling2012},
\begin{equation}
H_{\mathrm{SOC}} = i \lambda_{\text{SOC}} \sum_{\langle\langle i,j \rangle\rangle} 
\nu_{ij} \, c^{\dagger}_{i} \sigma_{z} c_{j} + h.c.
\label{Eq:ISOC}
\end{equation}
where $\lambda_{\text{SOC}}$ is the strength of the SOC. 
It takes the opposite signs for the up and down spins, and the factor $\nu_{ij}$ can choose $\pm 1$ depending on the direction of the hopping within the lattice. Inclusion of the SOC term only modifies the diagonal interaction $t_d$: $t_d \pm i\lambda_{\text{SOC}}$ depending upon the spin and coupling direction, keeping the other hopping integrals intact. Note that this SOC behaves like an effective internal magnetic field with directions being opposite for the two spins. However, unlike the external magnetic field, it does not break the time-reversal symmetry, as up and down spin electrons experience opposite effective magnetic fields. Very recently, both the intrinsic and extrinsic spin-orbital altermagnetism have been reported on a square lattice~\cite{Wang2025b}.
\begin{figure}
\centering
\includegraphics[scale=0.26]{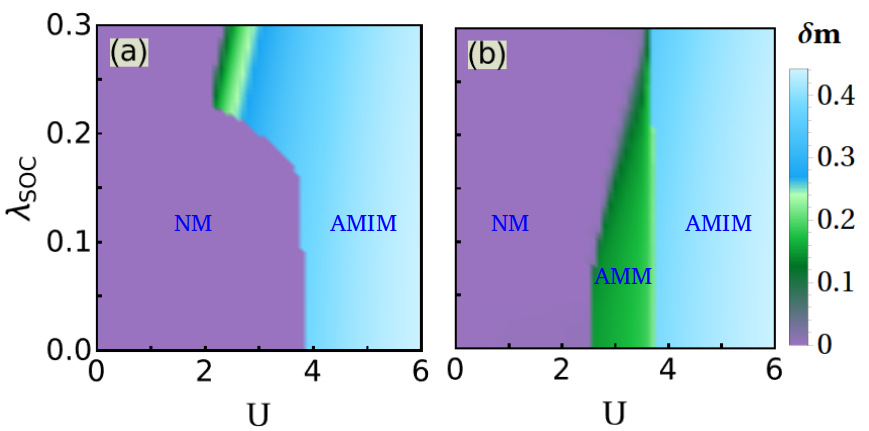}
\caption{Density profile of $\delta m$ over $U$ and SOC ($\lambda_{\text{soc}}$) for (a) $t_d = 0$ and (b) $t_d = 0.5$. The other parameters are the same as in Fig.~\ref{fig:EK}.}
\label{fig:Density:ISOC}
\end{figure}

In Fig.~\ref{fig:Density:ISOC}, we show the density plot for the AM order parameter over the space spanned by the Hubbard and the SOC strength. Previously, in the absence of the SOC, the NM phase is found in the phase diagram for both $t_d = 0$ and $t_d \ne 0$ when the Hubbard interaction is weak. With the inclusion of the SOC, the emergence of three phases: NM, AMM, and AMIM remains preserved qualitatively. The magnitude of $U_{\text{th}}$ decreases from $3.9$ to $2.6$ with the increase of the SOC strength for $t_d=0$ (see Fig.~\ref{fig:Density:ISOC}(a)). Note that the NM phase found in the presence of the SOC is also an isolated band type, since the bands are separated by energy gaps (see Appendices for details) with one isolated band appearing around the Fermi energy. Also, there is a tendency for the emergence of AM phase as an intermediate phase (see Fig.~\ref{fig:Density:ISOC}(a)) even without considering any diagonal hopping. In fact, this intermediate region becomes prominent when we include the diagonal hopping as found in Fig.~\ref{fig:Density:ISOC}(b). On top of these phenomena, the emergence of the AMIM remains preserved, indicating the stability of the AM phases in the presence of the SOC. 
\begin{figure}
\centering
\includegraphics[scale=0.26]{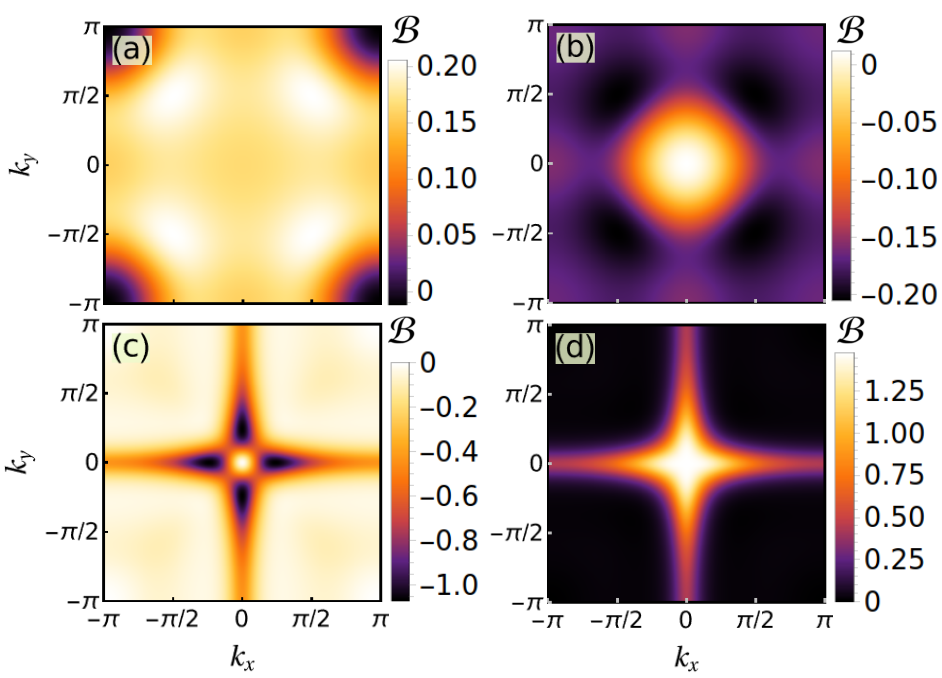}
\caption{Behavior of up-spin Berry curvature over the first BZ considering $\lambda_{\text{SOC}} = 0.14$, (a-b) $t_d = 0$ and (c-d) $t_d = 0.5$ for (a) $2$nd band ($\mathcal{C}^{\uparrow}_{2} = 1$), (b) $4$th band ($\mathcal{C}^{\uparrow}_{4} = -1$), (c) $3$rd band ($\mathcal{C}^{\uparrow}_{3} = -1$), and (d) $4$th band ($\mathcal{C}^{\uparrow}_{4} = 1$). The Hubbard interaction strength is taken as $U = 1.5$, and the other parameters are the same as in Fig.~\ref{fig:EK}.}
\label{fig:Berry}
\end{figure}

\subsection{Band Topology}

Finally, we discuss the band topology by calculating the Berry curvature and the Chern number of each band. The SOC of the lattice separates all energy bands from each other, which opens the possibility to examine the band topology. In the literature, the band topology has been explored previously in the absence of $U$~\cite{Bhattacharya2019}. We now compute the Berry curvature and Chern number associated with each band for finite $U$ and $t_d$, and show in Fig.~\ref{fig:Berry}. The Berry curvature associated with $n$-th band can be defined as,
\begin{widetext}
\begin{equation}
\mathcal{B}(E_{n},\mathbf{k}) = 
\sum_{E_{m} (\neq E_{n})} \dfrac{-2\, Im \Big[ \langle \psi_{n}(\mathbf{k}) | 
\Delta_{x} | \psi_{m}(\mathbf{k}) \rangle 
\langle \psi_{m}(\mathbf{k}) | 
\Delta_{y} | \psi_{n}(\mathbf{k}) \rangle \Big]} 
{(E_{n}-E_{m})^2},
\label{Eq:BC}
\end{equation}
\end{widetext}
where $\Delta_{x(y)} = \dfrac{\partial \bm{H_{tot}}(\mathbf{k})}{\partial k_{x(y)}}$ with $H_{tot}$ as the effective Hamiltonian including SOC and $\psi_{n}(k)$ as the eigenstate associated with energy $E_{n}$ for the $n$-th band. Additionally, the integration of Berry curvature over the first BZ gives the Chern number as,
\begin{equation}
\mathcal{C}_{n} = \frac{1}{2 \pi}\int_{-\pi}^{\pi}\mathcal{B}(E_{n},\mathbf{k})d{\mathbf{k}}.
\label{eq:Chern}
\end{equation}
Note that this Chern number is ill-defined when bands touch each other. Thus, the SOC is essential for Lieb-$5$ lattice for the present calculation.  
In the presence of the SOC, when no diagonal hopping is taken into consideration, the Chern number for each band turns out as $\mathcal{C}_{n}^{\uparrow (\downarrow)} = (0(0), +1(-1),0(0),-1(+1),0(0))$, where the band index is mentioned from the lower to higher energy, as marked in Fig.~\ref{fig:EKTI}(c). Thus, $2$nd and $4$th bands show a non-trivial topology, and the corresponding distribution of Berry curvature for topologically non-trivial spin-up bands is shown in Fig.~\ref{fig:Berry}(a,b). We show it only for one spin because of the degeneracy. Interestingly, as soon as we include the diagonal hopping, the picture of the band topology changes. Explicitly, $3$rd and $4$th bands exhibit non-trivial topological features, $\mathcal{C}_{n}^{\uparrow (\downarrow)} = (0(0), 0(0),-1(+1),+1(-1),0(0))$ (see Fig.~\ref{fig:EKTI}(d) for identifying the band index). Corresponding Berry curvature distributions are depicted in Fig.~\ref{fig:Berry}(c,d). So, under SOC without any diagonal hopping term $t_d = 0$, the phase transition happens from a topologically non-trivial phase to the AMIM phase, whereas the system permits a phase transition from the topological non-trivial phase to the AMIM phase by crossing an AMM phase for $t_d \ne 0$.

\section{Summary and conclusion} 
To summarize, we have shown the emergence of spin-lattice symmetric AM phases characterized by the sublattice degree of freedom, specifically, AMM and AMIM, in a repulsive Hubbard model on Lieb-$5$ lattice. By investigating the spin-resolved band structure, the altermagnetic order parameter, and the spectral function, we demonstrate the phase transition from the NM to either the AMIM or from the NM to AMIM via AMM phase through the phase diagram over a wide parameter space. The classification of the AM phases into the AMM and AMIM phases has been confirmed based on the appearance of the spin-split band structures and the amplitude of the order parameter. The emergence of AMIM and AMM orders in Lieb-5 lattices are energetically favored as they host the lowest ground state energy compared to FM or AFM orders. Note that, the AMM and AMIM phases appearing for the higher $U$ can be verified by using other approaches suitable for exact calculations, which is kept as a future outlook of the present work.

An AM phase with stability against the SOC has been identified without any diagonal hopping within the Lieb-5 lattice. The inclusion of the SOC also lowers the threshold values of the Hubbard strength. The unique AMIM phase, characterized by a single band crossing at the Fermi level and thus resulting in highly suppressed interband scatterings, has made our model promising for applications in quantum device components, particularly for switching operations. 

The experimental realization of the Lieb-$5$ lattice in Ga-doped YIG layer systems~\cite{Centala2023}, two-dimensional polymers composed of zinc-phthalocyanine (ZnPc)~\cite{Zhang2024c}, photonics~\cite{Hanafi2022,Zhang2017} enhances the potential of our proposal. Moreover, the observation of AFM orders in lattice Hubbard models~\cite{Hart2015,Mazurenko2017,Jordens2008} and the possibility of implementation of tunability of $U$ via Feshbach resonances~\cite{Chin2010} further strengthen the potential of our model for the realization of AM phases. 

\section{Acknowledgement}
We acknowledge the Department of Space (DoS), Government of India, for all support at Physical Research Laboratory (PRL) and also the ParamVikram-1000 HPC facility at PRL, Ahmedabad, for carrying out a major part of the calculations. S.\,B.\, thanks Arunava Chakrabarti and  Muktish Acharyya for some helpful discussions, and Purnendu Das for helpful communication.

\appendix

\section{Calculation of altermagnetic order parameter}
\label{AppendixA}

In this section, we present some additional discussions on calculations and some results to support our findings presented in the main text.

\textit{Momentum-space calculation:} In the main text, we show the altermagnetic order parameter ($\delta m$), which is solved using the Hartree-Fock equation self-consistently with a tolerance factor of $10^{-5}$,
\begin{figure}[ht!]
\centering
\includegraphics[scale=0.3]{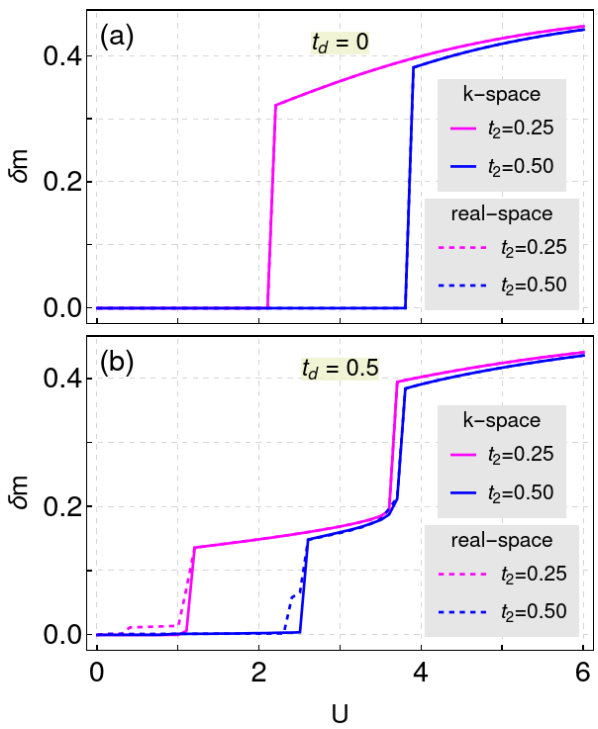}
\caption{Variation of $\delta m$ with the Hubbard interaction strength $U$ for an Lieb-$5$ lattice with (a) $t_d = 0$ and (b) $t_d = 0.5$. This calculation is performed in both the $k$-space and real-space ($N = 1620$) for two different values of intercell hopping strength $t_2$. Other parameters are kept the same as in Fig.~\ref{fig:EK}.}
    \label{fig:deltam-vs-U}
\end{figure}

\begin{figure}[ht!]
\centering
\includegraphics[scale=0.3]{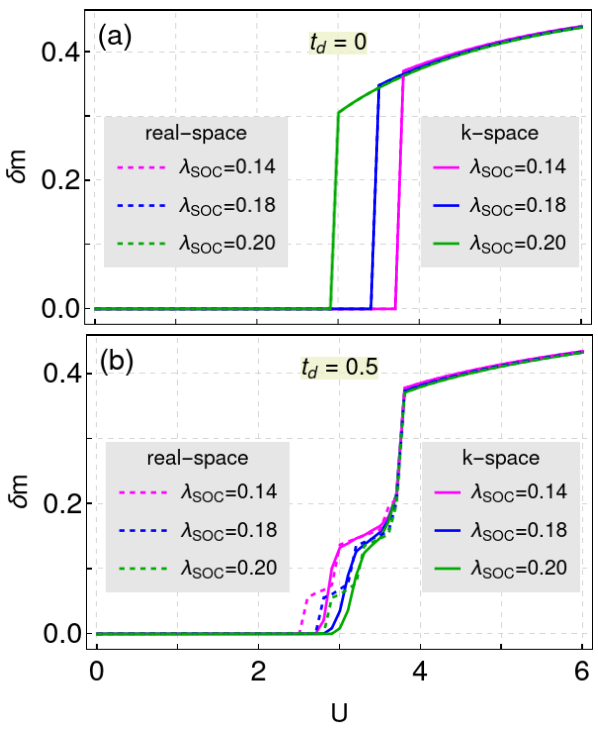}
\caption{Behavior of $\delta m$ with the Hubbard interaction strength $U$ in the presence of SOC, $\lambda_{\text{SOC}}$ for a Lieb-$5$ lattice with $t_2 = 0.5$ and (a) $t_d = 0$ and (b) $t_d = 0.5$. This calculation
is performed in both the k-space and real-space (N = 1620). Other parameters are the same as in Fig.~\ref{fig:EK}.}
\label{fig:deltam-vs-U-ISOC}
\end{figure}
\begin{equation}
\begin{split}
     \delta m & =  \frac{1}{8N} \sum_{k} \langle \sum_{i \in \alpha }(n_{k i \uparrow} - n_{k i \downarrow}) -\sum_{i \in \beta}(n_{ki \uparrow}-n_{ki \downarrow})\rangle_{HF} \\\
     &=\frac{1}{8 N} \sum_{k,j=1}^{10} \psi^{\dagger}_{j}(k)~ \Lambda ~\psi_{j}(k) \cdot f(E_{k,j} - \mu)
\end{split}
\label{Eq:finiteT}
\end{equation}
where $\Lambda$ is diag$[1,-1,1,-1,0,-1,1,-1,1,0]$. $E_{k,j}$, and $\psi_{j}(k)$ are the $j^{th}$ eigenvalues and eigenvectors of the Hamiltonian, and $f(E_{k,j})$ is the Fermi-Dirac distribution function. The chemical potential $\mu$ maintains the total fermionic density of the system: $n = \frac{1}{5 N} \sum_{k,j} f(E_{k,j} - \mu)$, where $N$ denotes the total number of $k_x, k_y$ points in the BZ. We maintain the half-filling condition ($n=1$) throughout the calculation. 

\textit{Real-space calculation:} In addition to the momentum space calculation of $\delta m$, we also repeat it in the real space using the same prescription. In the real space, the self-consistency equation of $\delta m$ takes the form,
\begin{equation}
\begin{aligned}
    \delta m
    &=\frac{1}{2N^{'}}\left(\sum_{i\in \alpha}\langle\ n_{i\uparrow}-n_{i\downarrow}\rangle
    -\sum_{i\in \beta}\langle n_{i\uparrow}- n_{i\downarrow}\rangle
    \right)\\
    &=\frac{1}{2N^{'}}\sum_{j=1}^{2N}
    \left[\sum_{i\in \alpha} \Delta_{i,j}^{\uparrow\downarrow} 
    \right.\left.
    -\sum_{i\in \beta}\Delta_{i,j}^{\uparrow\downarrow}
    \right]f(E_{j}-\mu)\\
\end{aligned}
\end{equation}
where $\Delta_{i,j}^{\uparrow \downarrow}$ = $(|\psi_{i,j \uparrow}|^2-|\psi_{i,j \downarrow}|^2)$ with $N$ as the total site index and $N^{'}$ indicating the total number of sub-lattice sites $\alpha$ and $\beta$. 
In Fig.~\ref{fig:deltam-vs-U}, we compare the behaviors of $\delta m$ found from the momentum-space and real-space calculations without any SOC. In the absence of the diagonal hopping, the behavior of $\delta m$ showing the transition from the NM to AMIM phase in the momentum-space shows excellent agreement with the behavior in the real-space for the entire range of $U$ as seen in Fig.~\ref{fig:deltam-vs-U}(a). The difference between the transition points for different $t_2$ values corroborates the results shown in the main text. In the presence of the diagonal hopping,
the step-like feature of $\delta m$ with the Hubbard strength as depicted in Fig.~\ref{fig:deltam-vs-U}(b), indicating two consecutive phase transitions, NM to AMM and AMM to AMIM, shows very good agreement between the momentum-space and real-space calculations. We confirm that the very small discrepancies will be solved by increasing the real-space lattice size. However, due to computational limitations, we have restricted our calculations to the present system size.
\begin{figure}
\centering
\includegraphics[scale=0.24]{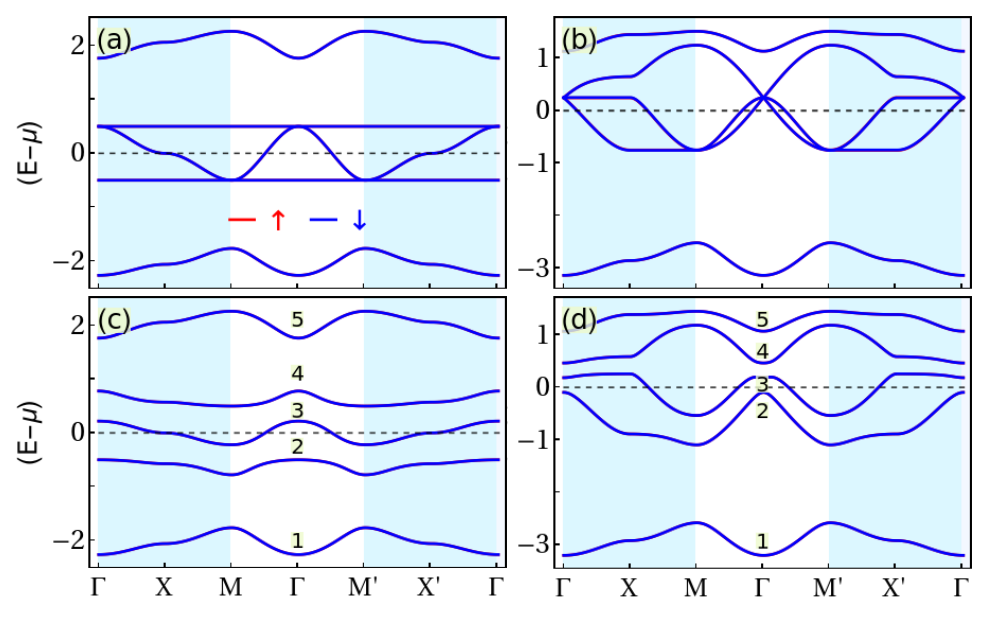}
\caption{(a,b) Spin-resolved band structure of Lieb-$5$ lattice (a-b) without and (c-d) with SOC. The diagonal hopping integral is taken as (a,c) $t_d = 0$ and (b,d) $t_d =0.5$. The strength of the SOC and Hubbard interaction is fixed as $\lambda_{\text{SOC}} = 0.14$ and $U = 1.5$. The other parameters are the same as in Fig.~\ref{fig:EK}.}
\label{fig:EKTI}
\end{figure}
We also repeat our calculations in the presence of the SOC and show in Fig.~\ref{fig:deltam-vs-U-ISOC} for various values of SOC. Similar to the previous case (without SOC), we found excellent agreements between the momentum-space and real-space calculations in the absence of the diagonal hopping, whereas there is a very slight discrepancy in the presence of the diagonal hopping, which will also decrease with the increase in the system size. 

In both Fig.~\ref{fig:deltam-vs-U} and Fig.~\ref{fig:deltam-vs-U-ISOC}, the zero $\delta m$ indicates the default NM phase of the present lattice, or equivalently, the absence of AM phase. $\delta m$ rises to a finite value when the AM phase appears. Now, the transitions from the NM to AMM and AMIM phases are further identified from the two-step behaviors of the AM order parameter. With the introduction of SOC, the threshold value of $U$, i.e., $U_{\text{th}}$, decreases with the increase in the SOC strength. Interestingly, the AMIM phase remains almost unaffected by the strong SOC.

\section{Band structures}
\label{AppendixB}
The AM band structures along the high symmetry paths are discussed in the main text. For completeness, we now discuss the full energy band structures and the total spectral density for different AM phases. 

\begin{figure}
\centering
\includegraphics[scale=0.35]{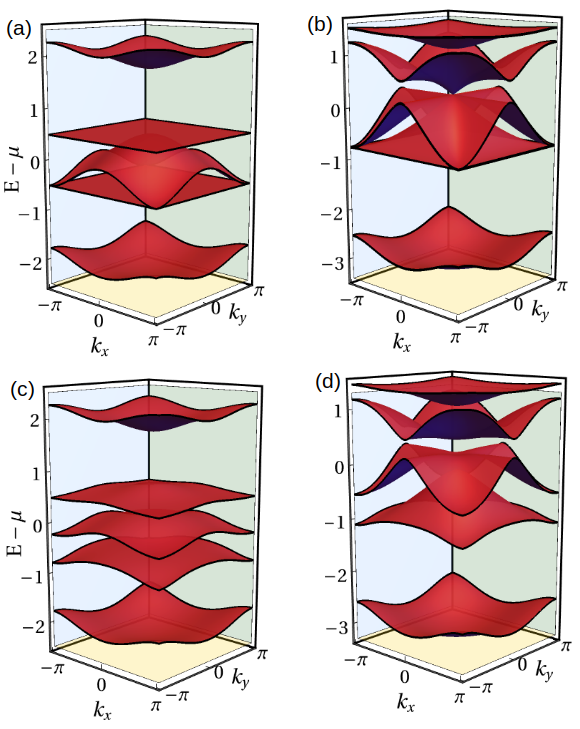}
\caption{Energy band dispersion of the Lieb-$5$ systems (a,b) without and (c,d) with inclusion of SOC, $\lambda_{\text{SOC}} = 0.14$. The diagonal hopping $t_d$ chosen for (a,c) $t_d = 0$ and for (b,d) $t_d = 0.5$. The Hubbard interaction strength is set at $U = 1.5$. Red (blue) colors are strands for spin up (down) bands. Other parameters remain fixed the same as in Fig.~\ref{fig:EK}.}
\label{fig:EK3D1}
\end{figure}
\begin{figure}
\centering
\includegraphics[scale=0.35]{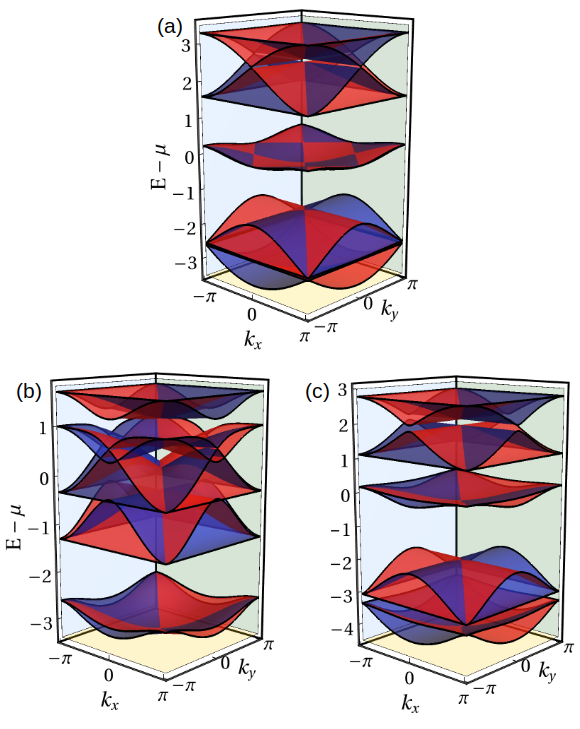}
\caption{Energy band structure of the Lieb-$5$ systems (a) without and (b,c) with the presence of diagonal hopping $t_d = 0.5$. The Hubbard interaction strength is set at (a,c) $U = 5$ and (b) $U = 3$. Red (blue) colors are strands for spin up (down) bands. Other parameters remain fixed the same as in Fig.~\ref{fig:EK}.}
\label{fig:EK3D2}
\end{figure}

In Fig.~\ref{fig:EKTI}, we present the band structure for a lower Hubbard strength. It shows the normal phases without any spin-splitting. Note that, in the absence of the SOC, a band touching is observed around the Fermi energy for both $t_d = 0$ (see Fig.~\ref{fig:EKTI}(a)) and $t_d = 0.5$ (see Fig.~\ref{fig:EKTI}(b)) regimes. Moreover, with the inclusion of the SOC, the bands get separated from each other, but still without any spin-splitting. Corresponding band structures around the high symmetry path are depicted in Fig.~\ref{fig:EKTI}(c,d). For a better understanding, we also compute the full band structure in the $k_x$, $k_y$ space as depicted in Fig.~\ref{fig:EK3D1} and confirm the appearance of spin-degenerate energy bands and the effect of SOC. The appearance of AM phases is already discussed in the main text, and corresponding spin-resolved bands along the symmetry path are shown in Fig.~\ref{fig:EK}. For completeness, we plot the corresponding spin-resolved band dispersion for the full BZ in Fig.~\ref{fig:EK3D2}. The momentum-dependent splitting of energy bands is apparent for both $t_d = 0$ (see Fig.~\ref{fig:EK3D2}(a)) and $t_d = 0.5$ (see Fig.~\ref{fig:EK3D2}(b,c)) with different Hubbard interaction strengths. Recently, these momentum-dependent energy bands have been realized in various materials using high-resolution ARPES~\cite{Li2025, Krempasky2024, Fedchenko2024, Lee2024, Yang2025}. 

\begin{figure}
\centering
\includegraphics[scale=0.26]{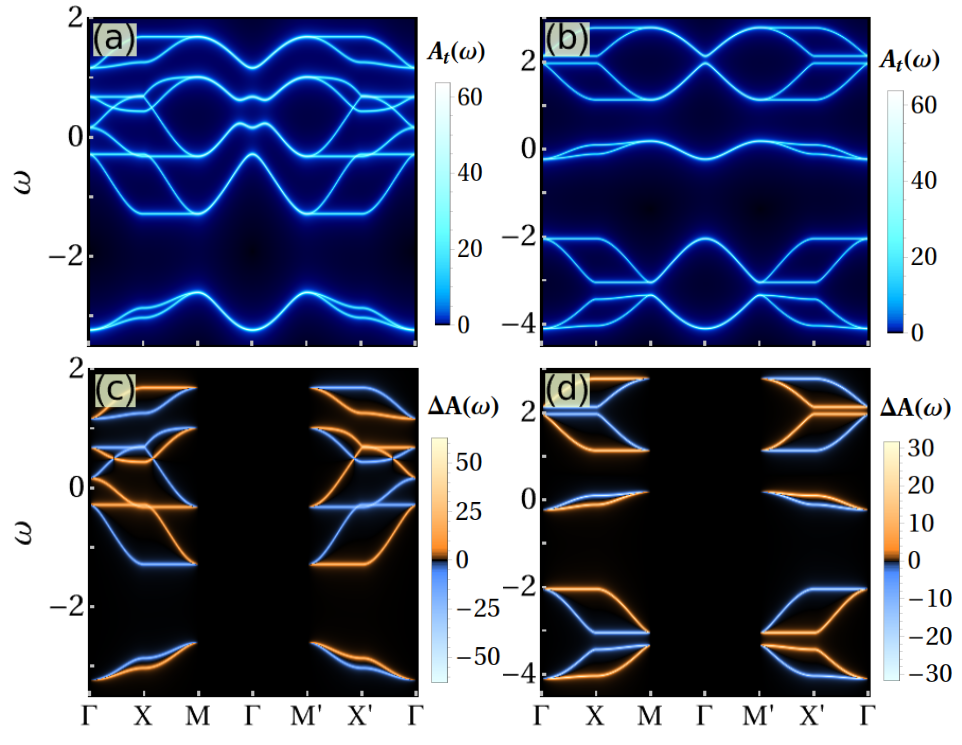}
\caption{Visualization of (a,b) total spectral density $A_{t}(\omega) = A_{\uparrow}(\omega)+A_{\downarrow}(\omega)$, (c,d) difference in spin-resloved spectral density $\Delta A(\omega) = A_{\uparrow}(\omega)-A_{\downarrow}(\omega)$ over a trial energy range $\omega$ along high symmetry path for the Lieb-$5$ system with (a,c) $U = 3$ and (b,d) $U = 5$. The other parameters are the same as in Fig.~\ref{fig:spectral}.}
\label{fig:ARPES}
\end{figure}

In addition to the band structure, the behavior of the spectral function is also shown in the main text, but in terms of the difference between the spin components. We calculate the total spectral density $A_{t} (\omega) = A_{\uparrow}(\omega) + A_{\downarrow}(\omega)$ along the high-symmetry paths for a fixed energy range $\omega$ using Eq.~\ref{Eq:Spectral-function} and display in Fig~\ref{fig:ARPES}. Fig.~\ref{fig:ARPES}(a) and (b) are shown for different $U$ strengths, which indicate the AMM and AMIM phase, respectively. In Fig.~\ref{fig:ARPES}(c,d), we compute the difference in spin-resolved spectral density $\Delta A(\omega) = A_{\uparrow}(\omega) - A_{\downarrow}(\omega)$ along the symmetry path. When $\omega$ is very close to the own energy of the system, $A_{t}(\omega)$ appears with a higher amplitude and vanishing values otherwise. Due to this feature, $A_{t}(\omega)$ creates a bright intensity over the symmetry path around the location of the system's energy, showing a fantastic agreement with the energy band diagram. Noteworthy, the spin-polarization around the $\Gamma-X-M$ (or equivalently $M^{'}-X^{'}-\Gamma$) path is observed with a momentum inversion, whereas $\Delta A(\omega)$ is always zero around the $M-\Gamma-M^{'}$, indicating no spin-polarization. 

\section{Effect of Temperature and filling}
\label{AppendixC}
\begin{figure}
\centering
\includegraphics[scale=0.27]{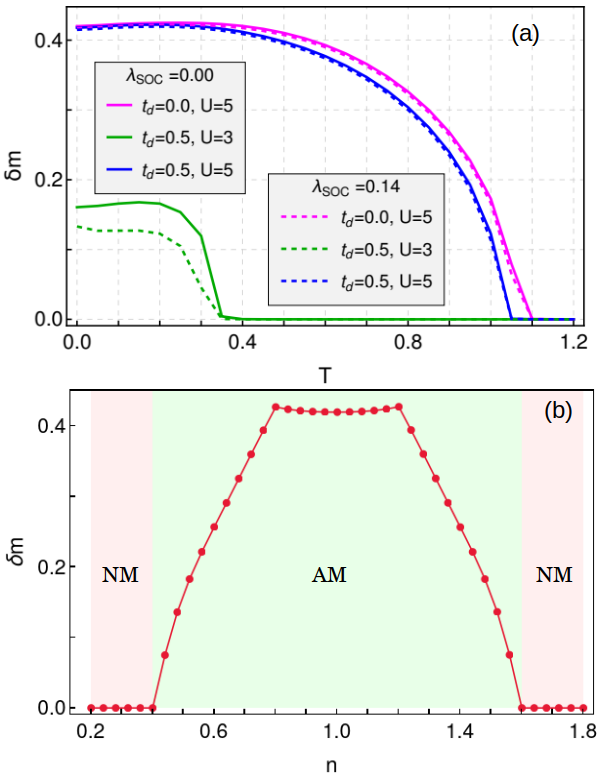}
\caption{Variation of AM order parameter with (a) temperature for without and with SOC, at $t_2 = 0.5$ and (b) electron filling with $t_2 = 0.25$, $t_d = 0$ and $U = 2.5$. The rest of the parameter values are taken the same as in Fig.~\ref{fig:EK}.} 
\label{fig:Temperature}
\end{figure}

Till now, we have considered the temperature $T = 0$ for all the results. From a realistic point of view, the stability of AM phases at finite temperatures is an important question to be addressed. The quantity temperature $T$ enters the calculation of the order parameter through the Fermi-Dirac distribution shown in Eq.~\eqref{Eq:finiteT}.
Figure~\ref{fig:Temperature}(a) shows the temperature dependence of the $\delta m$ for various parameter values both in the absence and presence of the SOC. Let us first discuss the scenario without any SOC (solid curves of Fig.~\ref{fig:Temperature}(a)). In the absence of the diagonal hopping, the emergence of the AMIM phase is only allowed. Now, with increasing temperature, the AM order parameter $\delta m$ gradually decreases and goes to zero at a critical temperature beyond which the system loses its AM character. Moreover, when the diagonal term is introduced, two AM phases appear. Among these, the lower $\delta m$ corresponding to the AMM phase also decreases with temperature and goes to zero at different $T_c$ compared to the other phase transition. The AMIM phase with a higher amplitude of $\delta m$ also follows a similar behavior, showing a higher $T_c$ compared to the AMM phase. To further validate our results, we compute the temperature dependence of $\delta m$ for all phases after including SOC (dotted lines of Fig.~\ref{fig:Temperature}). The nature of $\delta m$ with temperature is largely unaffected for AMIM phases. Under SOC, the amplitude of $\delta m$ for the AMM phase contains lower values compared to the AMM phase without SOC. However, their $T_c$ remains the same. 

In the main text, we show all results for half-filling where the total fermionic density per site of the system is unity maintained by the chemical potential $\mu$. However, to understand the behavior of AM phases beyond half-filling, we compute the altermagnetic order parameter for different fillings and show it in Fig.~\ref{fig:Temperature}(b). We observe that near half-filling, the alternant magnetic nature is preserved. However, the AM phase can be metallic or insulating depending on the exact value of the filling. For low or high filling, the spin splitting vanishes. It is not surprising because at lower filling, maximum sites are empty, not possible to maintain the required up and down spin distribution to get the altermagnetism. Moreover, when filling is higher, there is a tendency to the double occupancy of the electrons, which destroys the required arrangement to get the AM phase.

\section{Other traditional magnetic orders}
\label{AppendixD}
\begin{figure}
\centering
\includegraphics[width=\linewidth]{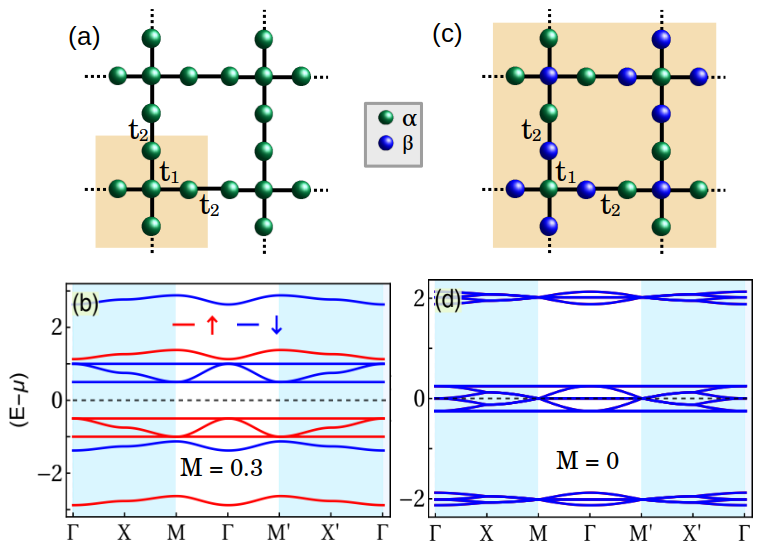}
\caption{(a) Ferromagnetic and (c) Anti-ferromagnetic ordering in Lieb-$5$ lattice. The yellow shaded box indicates the unit cells. (b,d) Spin-resolved band structures at (b) Ferromagnetic and (d) Anti-Ferromagnetic state with a Hubbard strength $U = 2.5$. The hopping $t_2$ and $t_d$ are chosen as $0.25$ and zero. The other parameters are the same as in Fig.~\ref{fig:EK}. The magnetization is denoted by $M$.}
\label{fig:FM_AFM}
\end{figure}

Although our main focus is on the AM phases, we now discuss the possibility of traditional magnetic phases in the Lieb-$5$ Hubbard model. To observe an FM order, a single sublattice is sufficient, as shown in Fig.~\ref{fig:FM_AFM}(a), unlike the model discussed in the main text. At half-filling, the occupation becomes $\langle n_{\sigma}\rangle = \frac{1}{2}+M(-1)^{\sigma}$, where $\sigma$ is $0$ (1) for $\uparrow (\downarrow)$ spin. The magnetization $M$ is given by, $M = \frac{1}{10 N} \sum_i \langle n_{i, \uparrow} - n_{i, \downarrow} \rangle$ where $N$ denotes the total number of $k_x$ $k_y$ points in the BZ. We self-consistently calculate the magnetization and the band structure and show in Fig.~\ref{fig:FM_AFM}(b). For this configuration, Lieb-5 lattice hosts a non-zero magnetization ($M = 0.3$) with a separation between the up and down energy bands.

To achieve an AFM order in this lattice, two magnetic sublattices $\alpha$ and $\beta$ are arranged within a unit cell as shown in Fig.~\ref{fig:FM_AFM}(c). The occupation becomes $\langle n_{\sigma} \rangle = \frac{1}{2} \pm M (-1)^{\sigma}$ for $\alpha$  ($\beta)$ sublattices and the system shows a zero magnetization ($M = 0$) without any band splitting as depicted in Fig.~\ref{fig:FM_AFM}(d) confirming the AFM order in the Lieb-$5$ systems. Note that the number of sites within a unit cell is higher in the AFM compared to the FM and AM counterparts.

Moreover, the emergence of a canted FM phase is also possible here. At the half filling, the occupation becomes: $\langle n \rangle = \frac{1}{2} + \vec{M}. \vec{\sigma}$ where $\vec{M} =(M_x,M_y,M_z)$ and $\sigma = (\sigma_x, \sigma_y, \sigma_z)$. To observe a canted FM phase, both longitudinal ($M_z$) and transverse ($M_x$) magnetization are needed, i.e., $\vec{M} = (M\sin{\theta}, 0, M\cos{\theta})$, where $M = \sqrt{M_x^2+M_z^2}$ is the magnitude of the magnetization and $\theta$ is the canting angle. Within the mean-field approximation, the Hubbard term of the Hamiltonian can be decoupled as: $-U M_z \sum_i(n_{i\uparrow}-n_{i\downarrow})-U M_x\sum_i(c_{i\uparrow}^{\dagger}c_{i\downarrow}+c_{i\downarrow}^{\dagger}c_{i\uparrow})$. Note that, for a given choice of the parameter, changing the canting angle only changes the direction of the magnetization while the total magnetization remains unaltered.

Given the possibilities of other traditional magnetic phases, AFM and FM, in Lieb-$5$ lattice with different sublattice arrangements, the question regarding the most favored state appears. It is observed that for a given parameter regime, the AMIM phase has the lowest energy both in the absence and presence of $t_d$, and thus the AMIM phase is the energetically most suitable ground state. For lower Hubbard strength, the AMM phase has the lowest energy compared to the AFM phase. Note that, for lower $U$, a direct comparison with the FM phase is not possible since the FM phase requires a higher threshold value of $U$.

\bibliography{reference}{}

\end{document}